\documentclass[prb,superscriptaddress,showpacs,floatfix,tightenlines,10pt,twocolumn]{revtex4}
\usepackage{amssymb}
\usepackage{amsmath}
\usepackage{graphicx}

\renewcommand{\Im}{\operatorname{Im}}
\begin{document}

\title{Phase transitions induced by a lateral superlattice potential in a
two-dimensional electron gas}
\author{Alice Blanchette}
\affiliation{D\'{e}partement de physique and Institut quantique, Universit\'{e} de
Sherbrooke, Sherbrooke, Qu\'{e}bec, J1K 2R1, Canada}
\author{Ren\'{e} C\^{o}t\'{e}}
\affiliation{D\'{e}partement de physique and Institut quantique, Universit\'{e} de
Sherbrooke, Sherbrooke, Qu\'{e}bec, J1K 2R1, Canada}

\begin{abstract}
We study the phase transitions induced by a lateral superlattice potential
(a metallic grid) placed on top of a two-dimensional electron gas (2DEG)
formed in a semiconductor quantum well. In a quantizing magnetic field and
at filling factor $\nu =1,$ the ground state of the 2DEG depends on the
strength $V_{g}$ of the superlattice potential as well as on the number of
flux quanta piercing the unit cell of the external potential. It was
recently shown\cite{Paper1} that in the case of a square lateral
superlattice, the potential modulates both the electronic and spin density
and in some range of $V_{g}$, the ground state is a two-sublattice spin
meron crystal where adjacent merons have the global phase of their spin
texture shifted by $\pi ,$\textit{\ }i.e. they are "antiferromagnetically"
ordered. In this work, we evaluate the importance of Landau-level mixing on
the phase diagram obtained previously for the square lattice\cite{Paper1}
and derive the phase diagram of the 2DEG modulated by a triangular
superlattice. When Landau level mixing is considered, we find in this case
that, in some range of $V_{g},$ the ground state is a three-sublattice spin
meron crystal where adjacent merons of the same vorticity have the global
phase of their spin texture rotated by $120^{\circ }$ with respect to one
another. This meron crystal is preceded in the phase diagram by another
meron lattice phase with a very different spin texture that does not appear,
at first glance, to resolve the spin frustration inherent to an
antiferromagnetic ordering on a triangular lattice.
\end{abstract}

\date{\today }
\maketitle

\section{INTRODUCTION}

The study of commensurability effects on the magnetoresistance and
magnetization of the two-dimensional electron solid in a perpendicular
magnetic field\cite{Reviewsuperlattice,Superlattice2} has been recently
revived by the observation of a Hofstadter butterfly spectrum in graphene on
top of Boron nitride\cite{Butterfly,Butterfly2,Wenchen,Tapash,Gumbs} and
also by the possibility of creating new artificial structure such as
artificial graphene\cite{Artificial}. One technique used to study these
effects in GaAs/AlGaAs quantum well is the patterning of a lateral
two-dimensional superlattice (or grid) on top of the semiconductor
heterostructure that hosts the two-dimensional electron gas (2DEG)\cite%
{Melinte}. The superlattice grid creates a periodic potential $V_{g}\left( 
\mathbf{r}\right) $ at the position of the electron gas that modulates the
electronic density. When the 2DEG is also subjected to a perpendicular
magnetic field, it is then characterized by two length scales: the lattice
constant of the superlattice potential and the magnetic length $\ell =\sqrt{%
\hslash /eB},$ where $B$ is the applied magnetic field.

The ground state of the interacting 2DEG in a GaAs/AlGaAs quantum well is
fully spin polarized at filling factor $\nu =1$ i.e. the 2DEG is a quantum
Hall ferromagnet\cite{Moon}. Its electronic density $n_{e}=1/2\pi \ell ^{2}$
is uniform and its Hall conductivity\cite{Hallreview} has the quantized
value $\sigma _{xy}=e^{2}/h$. In a previous work\cite{Paper1}, which we
shall refer to as Paper 1, one of us has studied the phase transitions
induced by a \textit{square} grid in the interacting 2DEG at $\nu =1$ in the
Hartree-Fock approximation. The phase diagram was studied for different
rational values of $\Gamma \equiv \varphi _{0}/Ba_{0}^{2}=2\pi \left( \ell
/a_{0}\right) ^{2}=q/p\in \left[ 0,1\right] ,$ where $\varphi _{0}=h/e$ is
the flux quantum, $a_{0}$ the lattice constant of the external grid and $q,p$
are integers with no common factors. The parameter $\Gamma ^{-1}$ represents
the number of flux quanta piercing a unit cell of the external potential. In
Paper 1, it was found that the ground state remains uniform and fully spin
polarized for finite $V_{g}$ up to a critical value $V_{g}^{\left( c\right)
} $ where a transition to a two-dimensional charge density wave (CDW)\ or
crystal takes place. Interestingly, this CDW is accompanied by a topological
spin texture that resembles that of a meron lattice with a two-sublattice
structure. As shown in Fig. 4 of Paper 1, the magnetic unit cell in this
particular CDW is twice the electronic unit cell and contains four spin
vortices (or merons)\ with the component $S_{z}$ of the spin being positive
at each vortex center. Each meron is surrounded by four neighboring merons
of opposite vorticity and so the topological charge alternates between $-1/2$
and $1/2$ from site to site, leading to positive and negative density
modulations of the uniform ground state. If we consider not just the
vorticity but the global phase of the spin vortex at each site, then the
four merons in a unit cell are divided into two pairs of merons with the
same vorticity but with global phases $0$ and $\pi $ (hence the name
"two-sublattice"). Treating this global phase as a spin, we may say that
merons with the same vorticity have an "antiferromagnetic" coupling. A
similar structure was found for the crystal of skyrmions\cite%
{SkyrmionsReview} that occurs \textit{near} filling factor $\nu =1$ in the
potential-free but interacting 2DEG\cite{Fertigskyrme}.

In Paper 1, it was assumed that the potential $V_{g}\left( \mathbf{r}\right) 
$ does not lead to Landau level mixing i.e. to occupation of the higher
Landau levels $n>0$. The calculation was done entirely within the two spin
levels of the $n=0$ Landau level. However, it is not a priori obvious that
this approximation is valid because a realistic value of $a_{0}$ leads to a
relatively small value of the magnetic field (see the next section where
this is discussed) thus possibly increasing the Landau-level mixing. It is
thus important to study the effect of Landau-level mixing on the phase
diagram found previously. We do this by adding level $n=1$ to the Hilbert
space. We then compute the occupation of Landau level $n=1$ as a function of 
$\Gamma $ and $V_{g}.$ Our results show that mixing is generally small
except at large values of $\Gamma $, but can lead to a qualitative change in
the phase diagram. For example, it modifies the phase below $V_{g}^{\left(
c\right) }$ for the square grid so that that the electronic density is no
longer uniform.

In this work, we also consider a triangular superlattice potential. Since
adjacent merons have their global phase rotated by $\pi $ (an
"antiferromagnetic ordering" to use a spin analogy), such a grid should lead
to frustration in the meron lattice. For a triangular lattice, it is well
known that this frustration is resolved by creating a three-sublattice
antiferromagnet where spins on adjacent sites are rotated by $120$ degrees.
We find that this is also true for the meron lattice: adjacent merons have
their global phase rotated by $120$ degrees thus creating a three-sublattice
spin meron crystal. We note that this type of structure does not occur if
the Hilbert space is restricted to the $n=0$ Landau level only. Just as its
bipartite counterpart on the square lattice\cite{Paper1}, we expect this
triangular meron lattice to sustain a gapless spin (Goldstone) mode\cite%
{CoteGirvin} while the phonon mode would be gapped by the external
potential. Surprisingly, we find that the three-sublattice phase is preceded
by another meron lattice phase with a different ordering of the global phase
of the meron that does not seem, at first glance, to resolve the frustration
inherent to an antiferromagnetic ordering on a triangular lattice.

Mixing of the $n=0$ and $n=1$ states can also be seen as introducing a
density of electric dipoles in the ground state. However, we find that all
phases in the phase diagram of the square or triangular lattice have a
texture of electric dipoles that is basically that imposed by the external
potential so that the different phases are not distinguishable from this
feature alone.

Our paper is organized as follows. In Sec. II, we introduce the superlattice
potential and the model parameters. In Sec. III, we briefly review the
Hartree-Fock approximation that we use to derive the phase diagram of the
interacting 2DEG. In Secs. IV and V, we present the phase diagram of the
square and triangular lattices respectively. We discuss the induced electric
dipole texture in Sec. VI and conclude in Sec. VII.

\section{SUPERLATTICE POTENTIAL AND\ MODEL\ PARAMETERS}

We consider a square or triangular lateral superlattice (grid) with a
lattice constant $a_{0}$ and a unit cell area $s=\varepsilon a_{0}^{2}.$ The
grid is placed on top of a GaAs/AlGaAs quantum well semiconductor
heterostructure. For the square(triangular) grid, $\varepsilon =1\left( 
\sqrt{3}/2\right) $. A transverse magnetic field, $\mathbf{B}=B\widehat{%
\mathbf{z}}$ is applied to the 2DEG and we define the parameter%
\begin{equation}
\Gamma =\frac{q}{p}=\frac{\varphi _{0}}{Bs},
\end{equation}%
where $q,p$ are integers with no common factors and $\varphi _{0}=h/e$ is
the flux quantum. The parameter $\Gamma ^{-1}$ is the number of flux quanta
piercing one unit cell of the external superlattice.

We consider the following simple form for the grid potential at the position
of the 2DEG 
\begin{equation}
V_{g}\left( \mathbf{r}\right) =\frac{1}{S}\sum_{\mathbf{G}_{0}}V_{g}e^{i%
\mathbf{G}_{0}\cdot \mathbf{r}},  \label{grid}
\end{equation}%
where $\mathbf{r}$ is a vector in the plane of the 2DEG and $\left\vert 
\mathbf{G}_{0}\right\vert =2\pi /a_{0}$ (square lattice) or $\left\vert 
\mathbf{G}_{0}\right\vert =4\pi /\sqrt{3}a_{0}$ (triangular lattice) are the
4 (square lattice) or 6 (triangular lattice) reciprocal lattice vectors
(RLV's)\ on the first shell of RLV's of the superlattice potential. Note
that our calculation could be carried on with a different form for $%
V_{g}\left( \mathbf{r}\right) $ if we need a more realistic expression for
the grid potential or if the electrostatic confinement is achieved by a more
complex potential than a simple grid.

We study the phase diagram of the 2DEG at filling factor $\nu =1,$ for
discrete values of $\Gamma \in \left[ 0,1\right] $ and for a fixed value of $%
a_{0}$ which we take as $a_{0}=50$ nm, an experimentally accessible value%
\cite{Melinte}. With $a_{0}$ and $\Gamma $ fixed, the magnetic field is
given in Tesla by%
\begin{equation}
B\text{ [T]}=\frac{h}{es\Gamma }=\frac{4135.\,\allowbreak 7}{\Gamma
\varepsilon \left( a_{0}\text{ [nm]}\right) ^{2}}\text{.}
\end{equation}%
The condition $\nu =1$ for the filling factor forces the density to be given
by%
\begin{equation}
n_{e}\text{[10}^{11}\text{ cm}^{-2}\text{]}=\frac{B\text{ [T]}}{%
4.\,\allowbreak 14}.
\end{equation}%
For a GaAs/AlGaAs quantum well, the dielectric constant is $\kappa =12.9,$
the gyromagnetic factor $\left\vert g^{\ast }\right\vert =0.45$ and the
effective mass $m^{\ast }=0.067m_{e},$ where $m_{e}$ is the bare electronic
mass. The cyclotron, Coulomb and Zeeman energies are then given by:%
\begin{eqnarray}
E_{cyc} &=&\hslash \omega _{c}^{\ast }=\frac{\hslash eB}{m^{\ast }}%
=1.\,\allowbreak 73B\text{ [T] meV,} \\
\Delta _{Z} &=&g^{\ast }\mu _{B}B=0.02\,\allowbreak 6B\text{ [T] meV,} \\
E_{coul} &=&\frac{e^{2}}{\kappa \ell }=4.\,\allowbreak 36\text{ }\sqrt{B%
\text{ [T]}}\,\text{meV.}
\end{eqnarray}%
If we use $e^{2}/\kappa \ell $ as our units of energy, then%
\begin{equation}
\widetilde{\Delta }_{Z}=\frac{g^{\ast }\mu _{B}B}{e^{2}/\kappa \ell }%
=5.\,\allowbreak 99\times 10^{-3}\sqrt{B\text{ [T]}}
\end{equation}%
and%
\begin{equation}
\widetilde{E}_{cyc}=\frac{\hslash \omega _{c}^{\ast }}{e^{2}/\kappa \ell }%
=0.40\sqrt{B\text{ [T]}}.
\end{equation}%
The Landau-level mixing increases with the ratio $\widetilde{E}_{cyc}$.
Figure 1 shows how the density $n_{e}$, the magnetic field $B,$ and the
cyclotron energy $\widetilde{E}_{cyc}$ vary with $\Gamma $ for the square
and triangular crystals of electrons when $a_{0}=50$ nm and $\nu =1$. For
the range of $\Gamma $ considered, this figure shows that the density and
magnetic field should be accessible experimentally. Figure 1(b) suggests
that Landau level mixing may be important for a grid parameter $a_{0}=50$
nm. Mixing can be reduced by decreasing $a_{0},$ but the magnetic field then
rapidly rises to very high values (for example, with $a_{0}=20$ nm, the
magnetic field is $B=52$ T for $\Gamma =1/5$ and a square lattice).

\begin{figure}[tbph]
\includegraphics[scale=0.9]{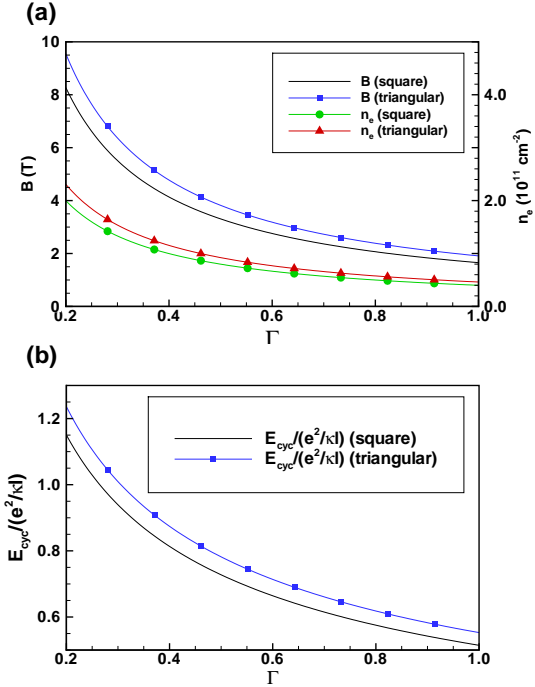}
\caption{(Color online)\ Behaviour of (a) the magnetic field $B$ and
electronic density $n_{e}$ and (b) the cyclotron energy ratio $%
E_{cyc}/(e^{2}/\protect\kappa \ell )$ as a function of the parameter $\Gamma 
$ for the square and triangular lattices at filling factor $\protect\nu =1$
and for a lattice constant of the external grid given by $a_{0}=50$ nm.}
\end{figure}

\section{HARTREE-FOCK HAMILTONIAN}

The Hartree-Fock Hamiltonian of the interacting 2DEG in the presence of the
grid is given by%
\begin{eqnarray}
H_{HF} &=&-N_{\varphi }\frac{\Delta _{Z}}{2}\sum_{\alpha
}\sum_{n}E_{n}^{\alpha }\rho _{n,n}^{\alpha ,\alpha }\left( 0\right)
\label{hamiltonien} \\
&&-\frac{eN_{\varphi }}{S}V_{g}\sum_{\mathbf{G}_{0}}\sum_{\alpha
}\sum_{n_{1},n_{2}}F_{n_{1},n_{2}}\left( -\mathbf{G}_{0}\right) \rho
_{n_{1},n_{2}}^{\alpha ,\alpha }\left( \mathbf{G}_{0}\right)  \notag \\
&&+N_{\varphi }\frac{e^{2}}{\kappa \ell }\sum_{\alpha ,\beta
}\sum_{n_{1},...,n_{4}}\sum_{\mathbf{G}\neq
0}H_{n_{1},n_{2},n_{3},n_{4}}\left( \mathbf{G}\right)  \notag \\
&&\times \left\langle \rho _{n_{1},n_{2}}^{\alpha ,\alpha }\left( -\mathbf{G}%
\right) \right\rangle \rho _{n_{3},n_{4}}^{\beta ,\beta }\left( \mathbf{G}%
\right)  \notag \\
&&-N_{\varphi }\frac{e^{2}}{\kappa \ell }\sum_{\alpha ,\beta
}\sum_{n_{1},...,n_{4}}\sum_{\mathbf{q}}X_{n_{1},n_{4},n_{3},n_{2}}\left( 
\mathbf{G}\right)  \notag \\
&&\times \left\langle \rho _{n_{1},n_{2}}^{\alpha ,\beta }\left( -\mathbf{G}%
\right) \right\rangle \rho _{n_{3},n_{4}}^{\beta ,\alpha }\left( \mathbf{G}%
\right) ,  \notag
\end{eqnarray}%
where $N_{\varphi }=S/2\pi \ell ^{2}$ is the Landau level degeneracy with $S$
the 2DEG area. The variables $\alpha ,\beta =\pm 1$ are spin indices while $%
n_{i}=0,1$ $\left( \text{with }i=1,2,3,4\right) $ are Landau level indices.
The $\mathbf{G}=0$ Fourier component in the Hartree term is cancelled by the
neutralizing positive background.

In Eq. (\ref{hamiltonien}), we have defined the operator 
\begin{eqnarray}
\rho _{n,m}^{\alpha ,\beta }\left( \mathbf{G}\right) &\equiv &\frac{1}{%
N_{\varphi }}\sum_{X,X^{\prime }}e^{-\frac{i}{2}G_{x}\left( X+X^{\prime
}\right) }  \label{un} \\
&&\times \delta _{X,X^{\prime }+G_{y}\ell ^{2}}c_{n,X,\alpha }^{\dagger
}c_{m,X^{\prime },\beta },  \notag
\end{eqnarray}%
where $c_{n,X,\alpha }^{\dagger }$ creates an electron with guiding-center
index $X$ (in the Landau gauge) with spin $\alpha $ in Landau level $n.$
These operators are related to the ground-state averaged electronic and spin
densities by%
\begin{eqnarray}
n_{\alpha }\left( \mathbf{r}\right) &=&\frac{1}{2\pi \ell ^{2}}%
\sum_{n_{1},n_{2}}\sum_{\mathbf{G}}F_{n_{1},n_{2}}\left( -\mathbf{G}\right)
\\
&&\times \left\langle \rho _{n_{1},n_{2}}^{\alpha ,\alpha }\left( \mathbf{G}%
\right) \right\rangle e^{i\mathbf{G}\cdot \mathbf{r}},  \notag \\
S_{\pm }\left( \mathbf{r}\right) &=&\frac{1}{2\pi \ell ^{2}}%
\sum_{n_{1},n_{2}}\sum_{\mathbf{G}}F_{n_{1},n_{2}}\left( -\mathbf{G}\right)
\\
&&\times \left\langle \rho _{n_{1},n_{2}}^{\pm ,\mp }\left( \mathbf{G}%
\right) \right\rangle e^{i\mathbf{G}\cdot \mathbf{r}},  \notag \\
S_{z}\left( \mathbf{r}\right) &=&\frac{\hslash }{2}\left[ n_{+}\left( 
\mathbf{r}\right) -n_{-}\left( \mathbf{r}\right) \right] ,
\end{eqnarray}%
where $S_{\pm }=S_{x}\pm iS_{y}$. The form factors $F_{n_{1},n_{2}}\left( 
\mathbf{G}\right) $ are defined by%
\begin{eqnarray}
F_{n_{1},n_{2}}\left( \mathbf{G}\right) &=&\sqrt{\frac{\min \left(
n_{1},n_{2}\right) !}{\max \left( n_{1},n_{2}\right) !}}L_{\min \left(
n_{1},n_{2}\right) }^{\left\vert n_{1}-n_{2}\right\vert }\left( \frac{%
G^{2}\ell ^{2}}{2}\right)  \label{trois} \\
&&\times e^{-G^{2}\ell ^{2}/4}\left( \frac{\left( \chi
_{n_{1},n_{2}}G_{y}+iG_{x}\right) \ell }{\sqrt{2}}\right) ^{\left\vert
n_{1}-n_{2}\right\vert },  \notag
\end{eqnarray}%
where $L_{n}^{m}\left( x\right) $ is a generalized Laguerre polynomial and $%
\chi _{n_{1},n_{2}}=$sgn$\left( n_{1}-n_{2}\right) $ with sgn the signum
function. The dimensionless Hartree and Fock interactions are defined by%
\begin{eqnarray}
H_{n_{1},n_{2},n_{3},n_{4}}\left( \mathbf{G}\right) &=&\frac{1}{q\ell }%
F_{n_{1},n_{2}}\left( \mathbf{G}\right) F_{n_{3},n_{4}}\left( -\mathbf{G}%
\right) ,  \label{y1} \\
X_{n_{1},n_{2},n_{3},n_{4}}\left( \mathbf{G}\right) &=&\int \frac{d\mathbf{p}%
\ell ^{2}}{2\pi }\frac{1}{p\ell }F_{n_{1},n_{2}}\left( \mathbf{p}\right)
\label{y2} \\
&&\times F_{n_{3},n_{4}}\left( -\mathbf{p}\right) e^{-i\widehat{\mathbf{z}}%
\cdot \left( \mathbf{p}\times \mathbf{G}\right) \ell ^{2}}.  \notag
\end{eqnarray}

The averaged "densities" $\left\{ \left\langle \rho _{n,m}^{\alpha ,\beta
}\left( \mathbf{G}\right) \right\rangle \right\} $ are the order parameters
that describe the various CDW or crystal states. They are computed by
solving the Hartree-Fock equation for the single-particle Green's function $%
G_{\alpha ,\beta }\left( \mathbf{G,}\tau \right) $ defined by%
\begin{eqnarray}
G_{n,m}^{\alpha ,\beta }\left( \mathbf{G,}\tau \right) &=&\frac{1}{%
N_{\varphi }}\sum_{X,X^{\prime }}e^{-\frac{i}{2}G_{x}\left( X+X^{\prime
}\right) } \\
&&\times \delta _{X,X^{\prime }-G_{y}\ell ^{2}}G_{n,m}^{\alpha ,\beta
}\left( X,X^{\prime },\tau \right) ,  \notag
\end{eqnarray}%
where%
\begin{equation}
G_{n,m}^{\alpha ,\beta }\left( X,X^{\prime },\tau \right) =-\left\langle
Tc_{n,X,\alpha }\left( \tau \right) c_{m,X^{\prime },\beta }^{\dagger
}\left( 0\right) \right\rangle .
\end{equation}%
The order parameters are obtained using the relation 
\begin{equation}
\left\langle \rho _{n,m}^{\alpha ,\beta }\left( \mathbf{G}\right)
\right\rangle =G_{m,n}^{\beta ,\alpha }\left( \mathbf{G,}\tau =0^{-}\right) .
\end{equation}

The equation of motion for $G_{n,m}^{\alpha ,\beta }\left( \mathbf{G,}\omega
_{n}\right) =\int_{0}^{\beta \hslash }d\tau e^{i\omega _{n}\tau }G\left( 
\mathbf{G},\tau \right) ,$ where $\omega _{n}=\left( 2n+1\right) \pi /\beta
\hslash $ with $\beta =1/k_{B}T$ is a fermionic Matsubara frequency, is a
straightforward generalization of Eq. (14) of Paper 1 that includes two
Landau levels instead of one. It leads to 
\begin{eqnarray}
&&\left[ i\omega _{n}-\left( E_{m,\alpha }-\mu \right) /\hslash \right]
G_{m,m^{\prime }}^{\alpha ,\beta }\left( \mathbf{G},\omega _{n}\right)
\label{eqmotion} \\
&=&\delta _{\mathbf{G},0}\delta _{m,m^{\prime }}\delta _{\alpha ,\beta } 
\notag \\
&&-\frac{1}{\hslash S}\sum_{\mathbf{G}^{\prime }}\sum_{n}V_{g}\left( \mathbf{%
G}-\mathbf{G}^{\prime }\right) F_{m,n}\left( \mathbf{G}-\mathbf{G}^{\prime
}\right)  \notag \\
&&\times \gamma _{\mathbf{G},\mathbf{G}^{\prime }}G_{n,m^{\prime }}^{\alpha
,\beta }\left( \mathbf{G}^{\prime },\omega _{n}\right)  \notag \\
&&+\sum_{n}\sum_{\mathbf{G}^{\prime }\neq \mathbf{G}}U_{m,n}^{H}\left( 
\mathbf{G-G}^{\prime }\right) \gamma _{\mathbf{G},\mathbf{G}^{\prime
}}G_{n,m^{\prime }}^{\alpha ,\beta }\left( \mathbf{G}^{\prime },\omega
_{n}\right)  \notag \\
&&-\sum_{n}\sum_{\gamma }\sum_{\mathbf{G}^{\prime }}\widetilde{U}%
_{m,n}^{F,\alpha ,\gamma }\left( \mathbf{G-G}^{\prime }\right) \gamma _{%
\mathbf{G},\mathbf{G}^{\prime }}G_{n,m^{\prime }}^{\gamma ,\beta }\left( 
\mathbf{G}^{\prime },\omega _{n}\right) ,  \notag
\end{eqnarray}%
where $\gamma _{\mathbf{G},\mathbf{G}^{\prime }}=e^{-i\mathbf{G}\times 
\mathbf{G}^{\prime }\ell ^{2}/2}$ and $n,m,m^{\prime }=0,1.$ The Hartree and
Fock potentials are defined by%
\begin{eqnarray}
U_{m,n}^{H}\left( \mathbf{G}\right) &=&\left( \frac{e^{2}}{\hslash \kappa
\ell }\right) \sum_{\gamma }\sum_{n_{1},n_{2}}H_{n_{1},n_{2},m,n}\left( -%
\mathbf{G}\right)  \label{hpoten} \\
&&\times \left\langle \rho _{n_{1},n_{2}}^{\gamma ,\gamma }\left( \mathbf{G}%
\right) \right\rangle ,  \notag \\
\widetilde{U}_{m,n}^{F,\alpha ,\gamma }\left( \mathbf{G}\right) &=&\left( 
\frac{e^{2}}{\hslash \kappa \ell }\right)
\sum_{n_{1},n_{2}}X_{n_{1},n,m,n_{2}}\left( -\mathbf{G}\right)
\label{fpoten} \\
&&\times \left\langle \rho _{n_{1},n_{2}}^{\gamma ,\alpha }\left( \mathbf{G}%
\right) \right\rangle  \notag
\end{eqnarray}%
and $n_{1},n_{2}=0,1.$

As described in Paper 1, the Hartree-Fock approximation leads to a set of $%
N_{\mathbf{G}}$ coupled and self-consistent equations where $N_{\mathbf{G}}$
is the number of RLVs kept in the calculation. This set of equations is
solved numerically using an iterative method. Good convergence after $%
\approx 1000$ iterations is achieved by taking $N_{\mathbf{G}}\approx 600.$

Once the Green's function is known, the density of states per area, $g\left(
\omega \right) ,$ can be obtained from the relation 
\begin{equation}
g\left( \omega \right) =-\frac{1}{2\pi ^{2}\ell ^{2}}\sum_{\alpha ,n}\Im%
\left[ G_{n,n}^{R,\alpha ,\alpha }\left( \mathbf{G}=0,\omega \right) \right]
,  \label{states}
\end{equation}%
where $G_{n,n}^{R,\alpha ,\alpha }\left( \mathbf{G}=0,\omega \right) $ is
the retarded single-particle Green's function obtained by the analytical
continuation $G_{n,m}^{\alpha ,\beta }\left( \mathbf{G,}i\omega
_{n}\rightarrow \omega +i\delta \right) $ of the Matsubara Green's function.
The Hartree-Fock transport gap (at $T=0$ K) can be extracted directly from $%
g\left( \omega \right) $ since the Fermi level is fixed by the condition
that $\nu =1.$

Note that the Hartree-Fock approach described here forces the CDW or crystal
to be commensurate with the lattice potential. In Paper 1, however, we
showed that a grid with a square unit cell of area $a_{0}^{2}$ could induce
a crystal with a spin texture periodicity $\sqrt{2}a_{0}\times \sqrt{2}a_{0}$%
. In order to describe this state and keep the grid potential unchanged, we
take the RLVs in the Hartree-Fock Hamiltonian of Eq. (\ref{hamiltonien}) to
be given by: $\mathbf{G}=\frac{2\pi }{\sqrt{2}a_{0}}\left( n,m\right) $,
with $n,m=0,\pm 1,\pm 2,...$ and take $\mathbf{G}_{0}$ in Eq. (\ref{grid})
to be on the \textit{second} shell of these new RLV's thus ensuring that $%
\left\vert \mathbf{G}_{0}\right\vert =2\pi /a_{0}$ is unchanged. We use a
similar trick for the triangular grid, which, as we find in the present
work, can induce a crystal with a spin texture periodicity $\sqrt{3}%
a_{0}\times \sqrt{3}a_{0}.$ In this case, the grid potential is unchanged if
we take $\mathbf{G}_{0}$ to be on the \textit{third} shell of the new RLV's
given by $\mathbf{G}=n\mathbf{G}_{1}+m\mathbf{G}_{2}$ where $n,m=0,\pm 1,\pm
2,...$ and $\mathbf{G}_{1}=\frac{2\pi }{\sqrt{3}a_{o}}\left( \frac{2}{\sqrt{3%
}},0\right) ,$ $\mathbf{G}_{2}=\frac{2\pi }{\sqrt{3}a_{o}}\left( \frac{1}{%
\sqrt{3}},1\right) .$

\section{PHASE DIAGRAM OF THE\ 2DEG FOR\ THE\ SQUARE\ GRID}

We now derive the phase diagram of the 2DEG as a function of the potential
strength $V_{g}$ [see Eq. (\ref{hamiltonien})] for a square grid. We take $%
\nu =1$, $a_{0}=50$ nm and $\Gamma \in \left[ 0,1\right] $ including the two
Landau levels $n=0,1.$ The phase diagram that we find is shown in Fig. 2.
(Note that the lines connecting the $\Gamma $ points in this graph are
merely a guide to the eyes and not a true phase boundary.) The numbers above
the red symbols indicate the magnetic field $B$ in Tesla that corresponds to
the corresponding value of $\Gamma .$

The phase diagram contains four phases. The electronic density is modulated
spatially in all of them i.e. $n\left( \mathbf{r}\right) =n_{0}+\delta
n\left( \mathbf{r}\right) ,$ where $n_{0}=1/2\pi \ell ^{2}$ is the uniform
density of a filled Landau level and $\delta n\left( \mathbf{r}\right) $ is
the density modulation. The magnitude of the spin density is also modulated
spatially and, for some phases, the orientation of the spins as well. The
different phases are:

\begin{enumerate}
\item CDW1: a fully spin polarized CDW with all spins pointing in the
direction of the external magnetic field. Both $n\left( \mathbf{r}\right) $
and $S_{z}\left( \mathbf{r}\right) $ have the periodicity $a_{0}\times a_{0}$
of the external potential. CDW1 is the ground state for $V_{g}\lesssim 2.0$
meV at all values of $\Gamma .$

\item VORTEX1: a CDW with a $\sqrt{2}a_{0}\times \sqrt{2}a_{0}$ magnetic
unit cell$.$ The density modulation is accompanied by a topological spin
texture $\mathbf{S}\left( \mathbf{r}\right) .$ The unit cell contains four
spin vortices: two with a counterclockwise rotation (negative vorticity i.e. 
$n_{v}=-1$) and two with a clockwise rotation (positive vorticity i.e. $%
n_{v}=1$). The spin texture is similar to that shown in Fig. 4 of Paper 1
where only level $n=0$ was considered. The spin density $S_{z}\left( \mathbf{%
r}\right) $ is everywhere positive. Adjacent vortices with the same
vorticity have the global phase of their spin texture rotated by $\pi $ with
respect to one another. In a language where the global phase is mapped into
an $xy-$spin, this $\pi $ rotation can be seen as an antiferromagnetic
ordering of the vortex pair. The topological spin texture of this phase is
reminiscent of that of the Skyrme crystal that occurs \textit{near} $\nu =1$
in the absence of an external potential\cite{Fertigskyrme}. The
antiferromagnetic ordering keeps the spins as parallel as possible
everywhere in space thus minimizing the exchange energy. The exchange energy
is minimal when all spins are parallel, a situation realized in a quantum
Hall ferromagnet.

\item VORTEX2: a phase similar to VORTEX1 with the same antiferromagnetic
ordering but with the vorticity and sign of $S_{z}$ at each vortex core
inverted. This phase is absent of the phase diagram for $\Gamma \geq 1/2.$

\item CDW2: a phase\ similar to CDW1 but only partially spin polarized with $%
\left\langle S_{z}\right\rangle =\left\vert \frac{1}{2}-\frac{\Gamma }{%
\varepsilon }\right\vert .$ There is no spin texture in this phase. The
density modulations have larger amplitude than in CDW1 because of the
stronger external potential.
\end{enumerate}

The topological three-dimensional spin texture associated with each vortex
is similar to that of a meron. In a meron, the spin points up or down at the
core center and tilts away from the $\widehat{\mathbf{z}}$ direction away
from the core. At large distance from the core, the spins point purely
radially in the $x-y$ plane. For a meron core at $\mathbf{r}=0$ and for a
field of unit spins, the topological charge $Q$ of a meron is defined by%
\begin{equation}
Q=\frac{1}{2}\left[ S_{z}\left( \infty \right) -S_{z}\left( 0\right) \right]
n_{v},  \label{topo}
\end{equation}%
where $n_{v}$ is the vortex winding number (i.e. the number of $2\pi $
rotation around the vortex core)\cite{Moon}. There are four flavors of meron
and they all have half the topological charge of a skyrmion or antiskyrmion
i.e. $\left\vert Q\right\vert =\frac{1}{2}.$ In this work, we associate a
positive vorticity, $n_{v}=1,$ with a clockwise rotation of the spins. For $%
n_{v}=-1,$ a positive $S_{z}$ at the meron core is thus associated with a
topological charge $Q=+1/2$ and, by the spin-charge coupling, with a
positive density modulation (a local increase with respect to the uniform
density i.e. $\delta n\left( \mathbf{r}\right) >0$). The opposite is true
for $Q=-1/2$ i.e. $\delta n\left( \mathbf{r}\right) <0.$ Since no charge are
added to the 2DEG which is kept at $\nu =1,$ there is an equal number of
merons ($Q=+1/2$) and antimerons ($Q=-1/2$). We remark here that we use the
word "meron" in a loose sense since we are not really dealing with a
classical field of unit spins but rather with a spin field that can be
modulated both in orientation and in density. It follows that the charge in
our "merons" is not quantized. As $V_{g}$ is increased, $\delta n\left( 
\mathbf{r}\right) $ at the center of the vortices with positive density
modulation becomes larger than $\delta n\left( \mathbf{r}\right) $ at the
center of the vortex with negative modulation but $\int \delta n\left( 
\mathbf{r}\right) d\mathbf{r}=0.$

\begin{figure}[tbph]
\includegraphics[scale=0.9]{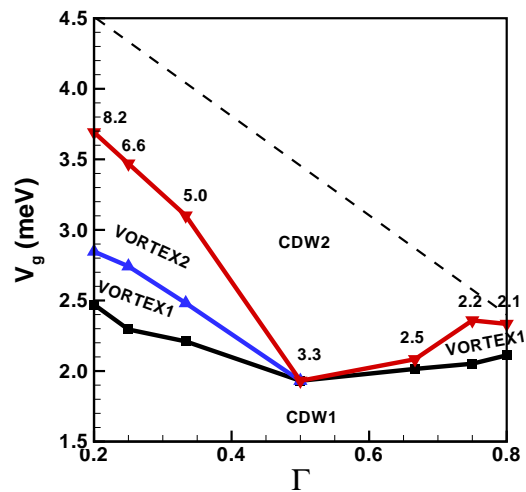}
\caption{(Color online) Phase diagram of the 2DEG under a square
superlattice potential $V_{g}$. The number above each red symbol gives the
magnetic field (in Tesla) for that particular value of $\Gamma .$ Note that
the lines are only a guide to the eye. Only the symbols correspond to
calculated values. The dashed line indicates the upper limit in $V_{g}$ of
our numerical calculation.}
\end{figure}

The addition of a second Landau level modifies the results reported in Paper
1. The most dramatic change is the disappearance of the uniform phase which
was present before the VORTEX1 phase and its replacement by the CDW1 phase
which now evolves continuously into the VORTEX1 state. The other two phase
boundaries (VORTEX1-VORTEX2 and VORTEX2-CDW2) are only slightly modified by
the addition of the second Landau level.

To evaluate the importance of the second Landau level, we compute the
occupation of the four Landau levels as a function of the grid potential $%
V_{g}$. Figure 3 shows these occupations for $\Gamma =3/4$ where, according
to Fig. 1, the Landau level mixing is expected to be the strongest. Our
calculation shows that the mixing is small, but not negligible at that value
of $\Gamma $. It varies very slightly with the grid potential in the range
of values considered. The presence of the second Landau level is especially
important at small value of $V_{g}$ where, as only $\nu _{0,\uparrow }$ and $%
\nu _{1,\uparrow }$ are nonzero, it allows for the formation of a non
uniform state with no spin texture. As $V_{g}$ increases, the occupation of
the $n=0,\alpha =\downarrow $ state more important than that of the $%
n=1,\alpha =\uparrow $ state. For $\Gamma <3/4,$ the occupation of the $n=1$
Landau level is much smaller than for $\Gamma =3/4.$ For example, it is $6\%$
at $\Gamma =1/2$ and $3.5\%$ at $\Gamma =1/4$ so that mixing is indeed small
for small $\Gamma ^{\prime }s.$

\begin{figure}[tbph]
\includegraphics[scale=0.9]{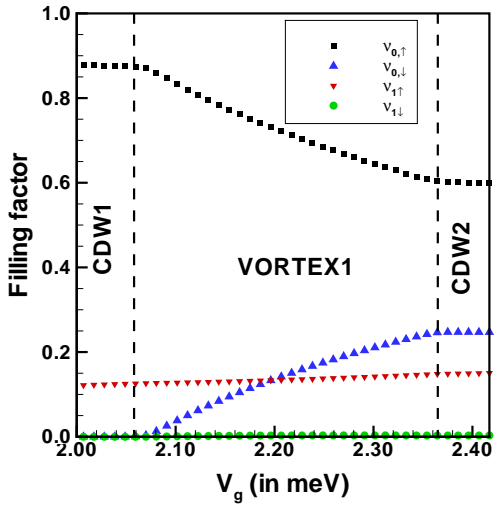}
\caption{(Color online) Occupation (filling factor) of Landau levels $n=0,1$
with spin $\protect\sigma =\pm 1$ as a function of the grid potential $V_{g}$
(in meV) for the square lattice at $\Gamma =3/4.$ The vertical dashed lines
indicate the phase boundaries.}
\end{figure}

In the square lattice, the transition between the VORTEX1 and VORTEX2 phases
for $\Gamma <1/2$ and between the VORTEX1 and CDW2 phases for $\Gamma >1/2$
is accompanied by a discontinuity in the spin component $S_{z}$ which we
show in Fig. 4 for different values of $\Gamma $. The discontinuity
increases with $\Gamma $ until it reaches a maximum (in absolute value) at $%
\Gamma =1/2$ where CDW1 is fully spin polarized while CDW2 is spin
unpolarized. We found in Paper 1 that $\Delta S_{z}=0$ for the VORTEX1-CDW2
transition but this is no longer the case when Landau level $n=1$ is
considered except when $\Gamma \geq 3/4.$ This discontinuity in $S_{z}$
should be detectable experimentally.

\begin{figure}[tbph]
\includegraphics[scale=0.9]{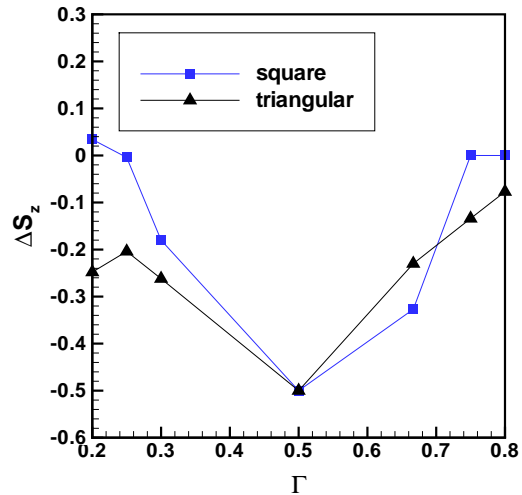}
\caption{(Color online) Discontinuity in the $z$ component of the spin
density as a function of $\Gamma $ for the square and triangular grids. For $%
\Gamma <1/2,$ the discontinuity is between the VORTEX1 and VORTEX2 phases
for the square lattice and VORTEX3 and VORTEX4 for the triangular lattice
while for $\Gamma >1/2$ it is between the VORTEX1 and CDW2 phases for the
square lattice and between CDW1 and CDW2 for the triangular lattice.}
\end{figure}

\section{PHASE DIAGRAM FOR\ THE\ TRIANGULAR\ GRID}

We now consider a triangular grid. The calculated phase diagram is shown in
Fig. 5. As in Fig. 2, the symbols correspond to calculated values and the
lines between them are guide to the eyes. The grid potential creates
modulated states with a triangular lattice structure. Two phases: CDW1 and
CDW2 are similar to the CDW's of the square potential. They occupy the
largest portion of the phase diagram. In between these two phases, we find
two vortex phases which we name VORTEX3 (see Fig. 6) and VORTEX4 (see Fig.
7). They are present for $\Gamma <1/2$ only. In contrast with the square
potential, they are not present if the Hilbert space is restricted to the
Landau level $n=0$ only (in which case we find only two phases: a uniform
phase, fully spin polarized and CDW2).

In the square lattice, we found that a pair of vortices (or merons)\ with
the same vorticity prefer an antiferromagnetic ordering of the phase of
their spin texture. In a triangular lattice, this type of interaction should
lead to frustration and the expected ground state must have a
three-sublattice structure with a $2\pi /3$ rotation of the phase from one
meron to the other (with the same $Q$). This is indeed what we find in the
VORTEX4 phase which is shown in Fig. 7. The magnetic unit cell is indicated
by the parallelogram. This phase has 3 merons of positive (negative)
vorticity at the each maximum (minimum) of the density. The spin component $%
S_{z}\left( \mathbf{r}\right) $ has not the same sign everywhere in space
but is negative at the core of each meron whatever its vorticity. Vortices
with $n_{v}=1$ thus correspond to density maxima (the red circles in Fig. 7)
according to Eq. (\ref{topo}) and those with $n_{v}=-1$ to density minima
(the large blue triangles in Fig. 7). The 3 merons with the same vorticity
have the phase of their spin texture rotated by $2\pi /3$ from one another.
They also have the same value of the local density modulation $\delta
n\left( \mathbf{r}\right) $ in contrast with the VORTEX3 phase discussed
below. Phase VORTEX4 is the analog of a three-sublattice antiferromagnet on
a triangular lattice. As for the square lattice, maximal and minimal values
of $\left\vert \delta n\left( \mathbf{r}\right) \right\vert $ are not equal
but depend on $V_{g}.$ However, $\int \delta n\left( \mathbf{r}\right) d%
\mathbf{r}=0$ as no charge is added to the 2DEG.

The second vortex phase that we find, VORTEX3, is shown in Fig. 6. The
magnetic unit cell is represented by the parallelogram. It contains 3
vortices with the same negative vorticity at the position of the density
maxima. The spin component (not shown in Fig. 6) $S_{z}\left( \mathbf{r}%
\right) $ is positive everywhere so that the 3 vortices are associated with
positive density modulations i.e. $\delta n\left( \mathbf{r}\right) >0$. The
negative modulation is spread throughout the unit cell and not concentrated
into antimerons. In contrast to VORTEX4, the 3 vortices in the unit cell do
not share the same value of $\delta n\left( \mathbf{r}\right) .$ One vortex
has a larger value of $\delta n\left( \mathbf{r}\right) $ than the other
two. Moreover, the spin texture of two of the three vortices have the same
global while the phase of the third one is shifted by $\pi $ with respect to
the other two. Considering the spin texture alone, VORTEX3 does not seem to
resolve the frustration inherent to an antiferromagnetic coupling on a
triangular lattice. We must keep in mind, however, that in our calculation
both the orientation and the spin can change locally i.e. $\left\vert 
\mathbf{S}\left( \mathbf{r}\right) \right\vert $ is not uniform. In Fig. 6,
vortices with the larger $\delta n\left( \mathbf{r}\right) $ are surrounded
by $6$ neighbors of opposite phase thus optimizing the antiferromagnetic
interaction. However, vortices with the smaller $\delta n\left( \mathbf{r}%
\right) $ are surrounded by 3 neighbors of opposite phase and 3 neighbors of
the same phase. For them, the interaction is not optimal. In total, however,
this seems to represent another way to resolve the frustration.

\begin{figure}[tbph]
\includegraphics[scale=0.9]{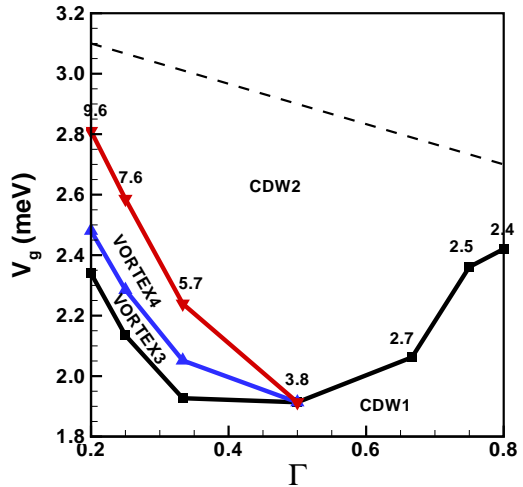}
\caption{(Color online) Phase diagram of the 2DEG under a triangular
superlattice potential $V_{g}$ (in meV). The number above each symbol gives
the magnetic field (in Tesla) for that particular value of $\Gamma .$ Note
that the lines connecting the symbols are only a guide to the eye. Only the
symbols correspond to calculated values. The dashed line indicates the limit
in $V_{g}$ of the numerical calculation.}
\end{figure}

\begin{figure}[tbph]
\includegraphics[scale=0.9]{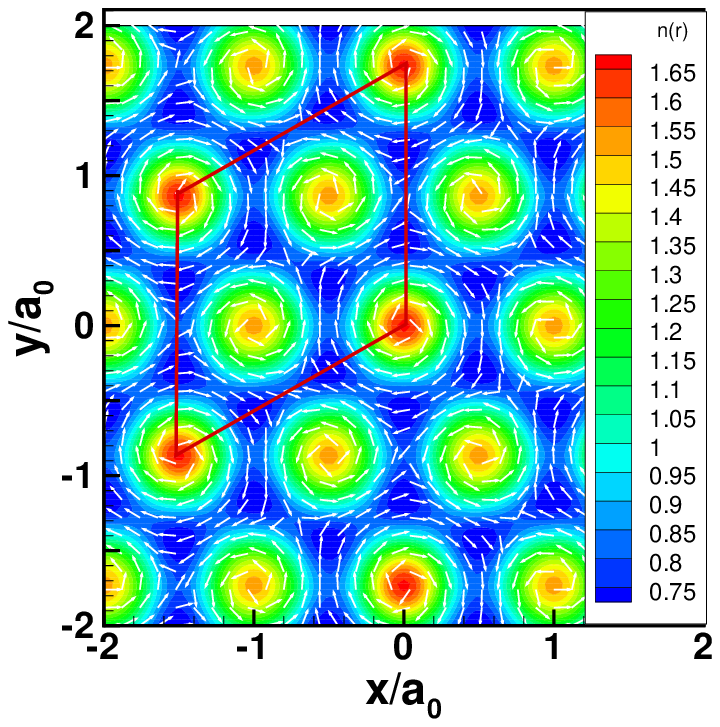}
\caption{(Color online) Electronic density $n\left( \mathbf{r}\right) $ in
units of $1/2\protect\pi \ell ^{2}$ and spin texture in the $x-y$ plane for
the VORTEX3 state. Parameters: $\Gamma =1/4,a_{0}=50$ nm and $\protect\nu %
=1. $ The magnetic unit cell is indicated by the parallelogram.}
\end{figure}

\begin{figure}[tbph]
\includegraphics[scale=0.9]{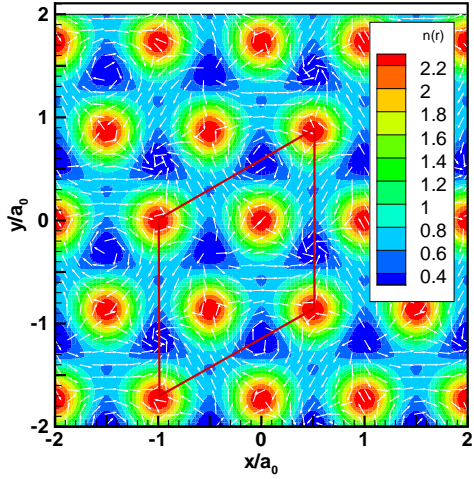}
\caption{(Color online) Electronic density $n\left( \mathbf{r}\right) $ in
units of $1/2\protect\pi \ell ^{2}$ and spin texture $\mathbf{S}\left( 
\mathbf{r}\right) $ in the $x-y $ plane for the VORTEX4 states. Parameters: $%
\Gamma =1/4,a_{0}=50$ nm and $\protect\nu =1.$ The magnetic unit cell is
indicated by the parallelogram. }
\end{figure}

We checked for the importance of Landau level mixing for the triangular
grid. Our results are shown in Fig. 8 for the case $\Gamma =1/3.$ Clearly,
the mixing is small at that value of $\Gamma $. But, for $\Gamma =4/5,$ the
occupation of the $n=1$ Landau level reaches $20\%$ and mixing becomes
significant as is the case for the square lattice. Figure 8 also clearly
shows the discontinuous character of the transition from the VORTEX3 to the
VORTEX4 phases.

\begin{figure}[tbph]
\includegraphics[scale=0.9]{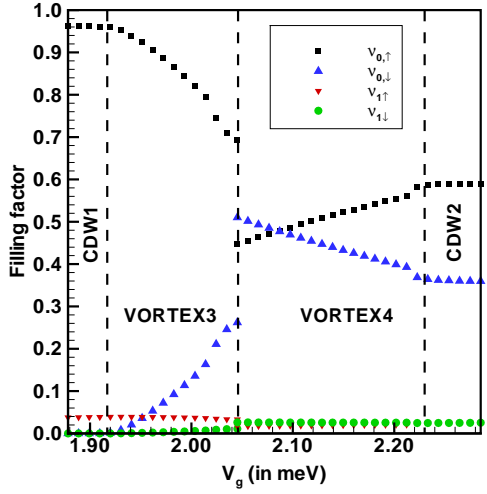}
\caption{(Color online) Filling factor $\protect\nu _{n,\protect\sigma }$ of
the four Landau levels $n=0,1$ with spin $\protect\sigma =\pm 1$ as a
function of the superlattice potential $V_{g}/\left( e^{2}/\protect\kappa %
\ell \right) $ for the triangular lattice at $\Gamma =1/3.$ The vertical
dashed lines indicate the boundaries of the different phases.}
\end{figure}

The transition between the VORTEX3 and VORTEX4 phases for $\Gamma <1/2$ or
between the CDW1 and CDW2 phases for $\Gamma >1/2$ is accompanied by a
discontinuity in $S_{z}$ which is plotted in Fig. 4. The behavior for the
triangular lattice is similar to that of the square lattice. The decrease in 
$S_{z}$ is maximal (in absolute value) for $\Gamma =1/2$ where the
transition is between the CDW1 (spin polarized) and CDW2 (spin unpolarized).

The VORTEX phases in the triangular lattice can also be distinguished from
the behavior of their transport gap $\Delta _{eh}$ which is the energy to
create an infinitely separated electron-hole pair$.$ We evaluate this gap
from the density of states given by Eq. (\ref{states}). Figure 9 shows the
dependency of $\Delta _{eh}$ on $V_{g}$ for $\Gamma =1/4.$ There is clear
change in the slope of $\Delta _{eh}$ between VORTEX3 and VORTEX4 and
between VORTEX4 and CDW2. The inset in Fig. 9 shows the density of states
and the integrated density of states (blue line) in the VORTEX3 phase at $%
V_{g}=2.17$ meV. The maximum value of the integrated DOS\ is $4$ in the
units of Fig. 9 corresponding to full occupation of the four states $n=0,1$
with $\sigma =\pm 1$. The gap is given by the size of the region where the
integrated DOS\ is unity.

\begin{figure}[tbph]
\includegraphics[scale=0.9]{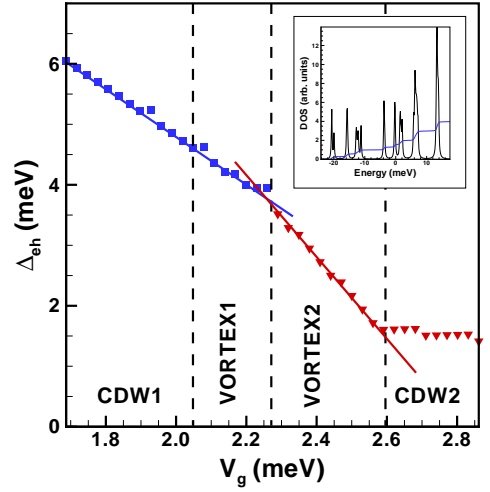}
\caption{(Color online) Behavior of the Hartree-Fock gap $\Delta _{eh}$ as a
function of the grid potential $V_{g}$ in the different phases of the
triangular lattice at $\Gamma =1/4.$ The gap is evaluated from the density
of states, an example of which is given in the inset for $V_{g}=2.17$ meV
i.e. in the VORTEX3 phase. The blue line in the inset is the integrated
density of states. }
\end{figure}

\section{ELECTRIC\ DIPOLES}

The averaged Fourier transform of the electronic density $n\left( \mathbf{G}%
\right) $ is related to the operator $\rho _{i,j}^{\alpha ,\beta }\left( 
\mathbf{G}\right) $ introduced in Eq. (\ref{un}) by the relation%
\begin{equation}
n\left( \mathbf{G}\right) =N_{\varphi }\sum_{\alpha }\sum_{n,m}F_{n,m}\left(
-\mathbf{G}\right) \left\langle \rho _{n,m}^{\alpha ,\beta }\left( \mathbf{G}%
\right) \right\rangle .
\end{equation}%
For $n,m=0,1,$ Eq. (\ref{trois}) gives for the form factors:%
\begin{eqnarray}
F_{0,0}\left( \mathbf{G}\right)  &=&e^{-G^{2}\ell ^{2}/4}, \\
F_{1,1}\left( \mathbf{G}\right)  &=&\left( 1-\frac{G^{2}\ell ^{2}}{2}\right)
e^{-G^{2}\ell ^{2}/4}, \\
F_{1,0}\left( \mathbf{G}\right)  &=&\left( \frac{\left( G_{y}+iG_{x}\right)
\ell }{\sqrt{2}}\right) e^{-G^{2}\ell ^{2}/4}, \\
F_{0,1}\left( \mathbf{G}\right)  &=&\left( \frac{\left( -G_{y}+iG_{x}\right)
\ell }{\sqrt{2}}\right) e^{-G^{2}\ell ^{2}/4}.
\end{eqnarray}%
If we use a pseudospin language where index $0$ is for orbital $n=0$ and $1$
for orbital $n=1,$ then the electronic and spin densities are given by 
\begin{eqnarray}
\rho ^{\alpha }\left( \mathbf{G}\right)  &=&\rho _{0,0}^{\alpha ,\alpha
}\left( \mathbf{G}\right) +\rho _{1,1}^{\alpha ,\alpha }\left( \mathbf{G}%
\right) , \\
\rho _{z}^{\alpha }\left( \mathbf{G}\right)  &=&\left( \rho _{0,0}^{\alpha
,\alpha }\left( \mathbf{G}\right) -\rho _{1,1}^{\alpha ,\alpha }\left( 
\mathbf{G}\right) \right) /2, \\
\rho _{0,1}^{\alpha ,\alpha }\left( \mathbf{G}\right)  &=&\rho _{x}^{\alpha
}\left( \mathbf{G}\right) +i\rho _{y}^{\alpha }\left( \mathbf{G}\right) , \\
\rho _{1,0}^{\alpha ,\alpha }\left( \mathbf{G}\right)  &=&\rho _{x}^{\alpha
}\left( \mathbf{G}\right) -i\rho _{y}^{\alpha }\left( \mathbf{G}\right) .
\end{eqnarray}%
It follows that%
\begin{eqnarray}
n\left( \mathbf{G}\right)  &=&N_{\varphi }\sum_{\alpha }\left( 1-\frac{%
G^{2}\ell ^{2}}{4}\right) \left\langle \rho ^{\alpha }\left( \mathbf{G}%
\right) \right\rangle e^{-G^{2}\ell ^{2}/4} \\
&&+N_{\varphi }\sum_{\alpha }\left( \frac{G^{2}\ell ^{2}}{2}\right)
\left\langle \rho _{z}^{\alpha }\left( \mathbf{G}\right) \right\rangle
e^{-G^{2}\ell ^{2}/4}  \notag \\
&&-N_{\varphi }\sqrt{2}i\sum_{\alpha }\left[ G_{x}\ell \left\langle \rho
_{x}^{\alpha }\left( \mathbf{G}\right) \right\rangle -G_{y}\ell \left\langle
\rho _{y}^{\alpha }\left( \mathbf{G}\right) \right\rangle \right]   \notag \\
&&\times e^{-G^{2}\ell ^{2}/4}.  \notag
\end{eqnarray}%
In this pseudospin language, the averaged coupling of the 2DEG with the grid
potential is written as%
\begin{eqnarray}
\left\langle W\right\rangle  &=&-\frac{eN_{\varphi }}{S}\sum_{\alpha
}\sum_{n,m}\sum_{\mathbf{G}}V_{e}\left( -\mathbf{G}\right) F_{n,m}\left( -%
\mathbf{G}\right)   \label{quatre} \\
&&\times \left\langle \rho _{n,m}^{\alpha ,\alpha }\left( \mathbf{G}\right)
\right\rangle   \notag \\
&=&-\frac{eN_{\varphi }}{S}\sum_{\alpha }\sum_{\mathbf{G}}V_{e}\left( -%
\mathbf{G}\right) \left( 1-\frac{G^{2}\ell ^{2}}{4}\right) \left\langle 
\overline{\rho }^{\alpha }\left( \mathbf{G}\right) \right\rangle   \notag \\
&&-\frac{eN_{\varphi }}{S}\sum_{\alpha }\sum_{\mathbf{G}}V_{e}\left( -%
\mathbf{G}\right) \left( \frac{q^{2}\ell ^{2}}{2}\right) \left\langle 
\overline{\rho }_{z}^{\alpha }\left( \mathbf{G}\right) \right\rangle   \notag
\\
&&+\frac{eN_{\varphi }}{S}\sqrt{2}i\sum_{\alpha }\sum_{\mathbf{G}%
}V_{e}\left( -\mathbf{G}\right)   \notag \\
&&\times \left[ q_{x}\ell \left\langle \overline{\rho }_{x}^{\alpha }\left( 
\mathbf{G}\right) \right\rangle -q_{y}\ell \left\langle \overline{\rho }%
_{y}^{\alpha }\left( \mathbf{G}\right) \right\rangle \right]   \notag
\end{eqnarray}%
where we have defined%
\begin{equation}
\overline{\rho }_{i}^{\alpha }\left( \mathbf{G}\right) =\rho _{i}^{\alpha
}\left( \mathbf{G}\right) e^{-G^{2}\ell ^{2}/4}.
\end{equation}%
The electric field $\mathbf{E}_{\Vert }\left( \mathbf{r}\right) $ in the
plane of the 2DEG is given by%
\begin{equation}
\mathbf{E}_{\Vert }\left( \mathbf{r}\right) =-\nabla V_{e}\left( \mathbf{r}%
\right) =-\frac{1}{S}\sum_{\mathbf{G}}i\mathbf{G}V_{g}\left( \mathbf{G}%
\right) e^{i\mathbf{G}\cdot \mathbf{r}},
\end{equation}%
and%
\begin{equation}
\mathbf{\nabla }_{\Vert }\cdot \mathbf{E}_{\Vert }\left( \mathbf{r}\right) =%
\frac{1}{S}\sum_{\mathbf{G}}G^{2}V_{g}\left( \mathbf{G}\right) e^{i\mathbf{G}%
\cdot \mathbf{r}}.
\end{equation}%
In real space, Eq. (\ref{quatre}) becomes 
\begin{eqnarray}
\left\langle W\right\rangle  &=&-eN_{\varphi }\sum_{\alpha }\int d\mathbf{r}%
\left\langle \overline{\rho }^{\alpha }\left( \mathbf{r}\right)
\right\rangle V_{g}\left( \mathbf{r}\right)  \\
&&+\frac{1}{2}\ell ^{2}eN_{\varphi }\sum_{\alpha }\int d\mathbf{r}%
\left\langle \overline{\rho }_{1,1}^{\alpha }\left( \mathbf{r}\right)
\right\rangle \mathbf{\nabla }_{\Vert }\cdot \mathbf{E}_{\Vert }\left( 
\mathbf{r}\right)   \notag \\
&&+\sqrt{2}\ell eN_{\varphi }\sum_{\alpha }  \notag \\
&&\times \int d\mathbf{r}\left[ \left\langle \overline{\rho }_{x}^{\alpha
}\left( \mathbf{r}\right) \right\rangle E_{x}\left( \mathbf{r}\right)
-\left\langle \overline{\rho }_{y}^{\alpha }\left( \mathbf{r}\right)
\right\rangle E_{y}\left( \mathbf{r}\right) \right] .  \notag
\end{eqnarray}%
The third line in this equation can be written as a coupling between a
dipole density and the electric field in the plane of the 2DEG, i.e.:%
\begin{equation}
\left\langle W\right\rangle _{dipole}=-\int d\mathbf{rd}\left( \mathbf{r}%
\right) \cdot \mathbf{E}_{\Vert }\left( \mathbf{r}\right) ,
\end{equation}%
if we define%
\begin{eqnarray}
\mathbf{d}\left( \mathbf{G}\right)  &=&-e\sqrt{2}\ell N_{\varphi
}\sum_{\alpha }\left( \left\langle \overline{\rho }_{x}^{\alpha }\left( 
\mathbf{G}\right) \right\rangle \widehat{\mathbf{x}}-\left\langle \overline{%
\rho }_{y}^{\alpha }\left( \mathbf{G}\right) \right\rangle \widehat{\mathbf{y%
}}\right) , \\
\mathbf{d}\left( \mathbf{r}\right)  &=&\frac{1}{S}\sum_{\mathbf{G}}\mathbf{d}%
\left( \mathbf{G}\right) e^{i\mathbf{G}\cdot \mathbf{r}}.
\end{eqnarray}%
The superposition of the $n=0$ and $n=1$ orbital states can thus be viewed
as creating a density of electric dipoles $\mathbf{d}\left( \mathbf{r}%
\right) $\cite{ShizuyaED}. Figure 10 shows the electronic density and dipole
texture for the VORTEX1 phase at $V_{g}/\left( e^{2}/\kappa \ell \right)
=0.22$ for the square lattice with $\Gamma =1/4.$ The dipole texture on each
lattice site is qualitatively the same for all phases (CDW's and VORTEX's).
In fact, the dipole field is basically that of the vector field $\mathbf{E}%
_{\Vert }\left( \mathbf{r}\right) $ in the plane of the 2DEG$.$ The electric
dipoles align themselves with the external potential of the grid. In
contrast with the spin field $\mathbf{S}\left( \mathbf{r}\right) $ which is
topologically different between the CDW's\ and VORTEX's phases, it does not
seem possible to distinguish between the different phases from their dipole
texture alone.

\begin{figure}[tbph]
\includegraphics[scale=0.9]{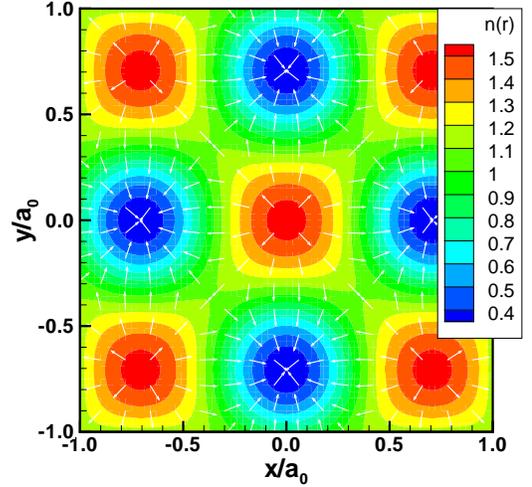}
\caption{(Color online) Electronic density $n\left( \mathbf{r}\right) $ and
dipole field $\mathbf{d}\left( \mathbf{r}\right) $ in the $x-y$ plane for
the VORTEX1 phase at $V_{g}/\left( e^{2}/\protect\kappa \ell \right) =0.22$
for the square lattice with parameters $a_{0}=50$ nm, $\Gamma =1/4$ and $%
\protect\nu =1.$}
\end{figure}

\section{CONCLUSION}

It appears from the numerical results that we have presented in this paper,
that the phase diagram of the 2DEG in a \textit{square} lateral superlattice
potential is not radically modified by the inclusion of a second Landau
level in the Hartree-Fock equation of motion. The main result is the
replacement of the uniform phase found in Paper 1 by a phase with density
modulation but no spin texture. As expected, the occupation of the $n=1$
Landau level increases with $\Gamma $ since a higher value of $\Gamma $
means a smaller magnetic field\textit{\ }i.e. more Landau level mixing.

For a triangular superlattice potential, a case not considered in Paper1,
the phase diagram is enriched by the introduction of the $n=1$ Landau level.
Apart from two CDW states with no spin modulation, we find vortex phases
analog to that found for the square superlattice. In particular, the VORTEX4
phase has the three-sublattice structure expected for a lattice of merons
where two adjacent merons with the same vorticity prefer to have the global
phase associated with their spin texture differing by a phase $\pi .$ As for
an antiferromagnet on a triangular lattice, the spin frustration that this
interaction create is resolved by having adjacent spins rotated by $120$
degrees leading to a three-sublattice antiferromagnet. The VORTEX3 phase
which occurs just before VORTEX4 seems to resolve the inherent frustration
of an antiferromagnetic ordering on a triangular lattice by having unequal
density for the three merons.

In Paper1, we calculated the collective mode dispersion of the different
phases of the square lattice in the generalized random-phase approximation
(GRPA). We showed that the VORTEX1 and VORTEX2 phases have an additional
Goldstone mode that is related to their spin texture, the Hartree-Fock
energy being independent of the global phase of the spin texture as is the
case in a Skyrme crystal\cite{CoteGirvin}. We expect that a similar
Goldstone mode should be present for the triangular superlattice, at least
for the VORTEX4 phase. Calculating the dispersion relation of the collective
modes is one way to ascertain the stability of a phase and it would be
interesting to be able to do it for the VORTEX3 state. When the $n=1$ Landau
level is considered, however, the size of the matrices involved in the GRPA\
calculation are of the order of $16N_{\mathbf{G}}\times 16N_{\mathbf{G}}$
where $N_{\mathbf{G}}\approx 600$ is the number of reciprocal lattice
vectors needed to described the vortex lattices. Those are too big matrices
to diagonalize with our current computational resources.

\begin{acknowledgments}
R. C. was supported by a grant from the Natural Sciences and Engineering
Research Council of Canada (NSERC). Computer time was provided by Calcul Qu%
\'{e}bec and Compute Canada.
\end{acknowledgments}


\begin{thebibliography}{99}
\bibitem{Paper1} R. C\^{o}t\'{e} and Xavier Bazier-Matte, Phys. Rev. B 
\textbf{94}, 205303 (2016).

\bibitem{Reviewsuperlattice} For an early review of this problem, see for
example: Daniela Pfannkuche and Rolf R. Gerhardts, Phys. Rev. B \textbf{46},
12606 (1992).

\bibitem{Superlattice2} A. Rauh, Phys. Status Solidi B \textbf{65}, K131
(1974); A. Rauh, Phys. Status Solidi B \textbf{69}, K9 (1975); R. R.
Gerhardts, D. Weiss, and K. v. Klitzing, Phys. Rev. Lett. \textbf{62}, 1173
(1989); R. W. Winkler, J. P.. Kotthaus and K. Ploog, Phys. Rev. Lett. 
\textbf{62}, 1177 (1989); Vidar Gudmundsson and Rolf R. Gerhardts, Phys.
Ref. B \textbf{52}, 16744 (1995).

\bibitem{Butterfly} D. R. Hofstadter, Phys. Rev. B \textbf{14}, 2239 (1976).

\bibitem{Butterfly2} B. Hunt, J. D. Sanchez-Yamagishi, A. F. Young, M.
Yankowitz, B. J. LeRoy, K. Watanabe,T. Taniguchi, P. Moon, M. Koshino, P.
Jarillo-Herrero, R. C. Ashoori, Science \textbf{340}, 1427 (2013); Till Schl%
\"{o}sser, Klaus Ensslin, J\"{o}rg P. Kotthaus and Martin Holland, Semicond.
Sci. Technol. \textbf{11}, 1582 (1996); C. R. Dean, L. Wang, P. Maher, C.
Forsythe, F. Ghahari, Y. Gao, J. Katoch, M. Ishigami, P. Moon, M. Koshino,
T. Taniguchi, K.Watanabe, K. L. Shepard, J. Hone and P. Kim, Nature \textbf{%
497}, 598 (2013); C. Albrecht, J. H. Smet, K. von Klitzing, D. Weiss, V.
Umansky, and H. Schweizer, Phys. Rev. Lett. \textbf{86}, 147 (2001); T.
Schloesser, K. Ensslin, J. P. Kotthaus, and M. Holland, Europhys. Lett. 
\textbf{33}, 683 (1996); Cheol-Hwan Park and Steven G. Louie, Nano Letters 
\textbf{9}, 1793 (2009); R. R. Gerhardts, D. Weiss, and U. Wulf, Phys. Rev.
B \textbf{43}, 5192 (1991).

\bibitem{Wenchen} Wenchen Luo and Tapash Chakraborty, J. Phys.: Condens.
Matter \textbf{28}, 0158801 (2016).

\bibitem{Tapash} V. M. Apalkov and T. Chakraborty, Phys. Rev. Lett. \textbf{%
112}, 176401 (2014); Areg Ghazaryan and Tapash Chakraborty, Phys. Rev. B 
\textbf{91}, 125131 (2015).

\bibitem{Gumbs} Godfrey Gumbs and Paula Fekete, Phys. Rev. B 56, 3787 (1997).

\bibitem{Artificial} Marco Gibertini, Achintya Singha, Vittorio Pellegrini,
Marco Polini, Giovanni Vignale and Aron Pinczuk, Phys. Rev. B \textbf{79},
241406(R) (2009).

\bibitem{Melinte} S. Melinte, Mona Berciu, Chenggang Zhou, E. Tutuc, S. J.
Papadakis, C. Harrison, E. P. De Poortere, Mingshaw Wu, P.M. Chaikin, M.
Shayegan, R.N. Bhatt, and R.A. Register, Phys. Rev. Lett. \textbf{92},
036802 (2004).

\bibitem{Moon} K. Moon, H. Mori, Kun Yang, S. M. Girvin, A. H. MacDonald, L.
Zheng, D. Yoshioka, and Shou-Cheng Zhang, Phys. Rev. B \textbf{51}, 5138
(1995).

\bibitem{Hallreview} For a review, see \textit{The quantum Hall Effect},
edited by R. E. Prange and S. M. Girvin (Springer-Verlag, New York, 1990)
and also the lecture notes of M. O. Goerbig, arXiv:0909.1998.

\bibitem{SkyrmionsReview} For a review on skyrmions, see Z. F. Ezawa, 
\textit{Quantum Hall Effects} (World Scientific, Singapore, 2000) or S. M.
Girvin and A. H. MacDonald in \textit{Perspectives in Quantum Hall Effects},
edited by S. Das Sarma and A. Pinczuk (Wiley, New York, 1997).

\bibitem{CoteGirvin} R. C\^{o}t\'{e}, H. A. MacDonald, Luis Brey, H. A.
Fertig, S. M. Girvin and H. T. C. Stoof, Phys. Rev. Lett. \textbf{78}, 4825
(1997).

\bibitem{Fertigskyrme} L. Brey, H. A. Fertig, R. C\^{o}t\'{e}, and A. H.
MacDonald, Phys. Rev. Lett. \textbf{75}, 2562 (1995).

\bibitem{ShizuyaED} K. Shizuya, Phys. Rev. B\textbf{79}, 165402 (2009).
\end{thebibliography}
\end{document}